\pdfoutput=1 
\documentclass[preprint,12pt]{elsarticle}
\newcommand{\be}{\begin{equation}}
\newcommand{\ee}{\end{equation}}
\newcommand{\ba}{\begin{eqnarray}}
\newcommand{\ea}{\end{eqnarray}}
\newcommand{\nn}{\nonumber}

\usepackage{amssymb}

\journal{.}

\begin{document}

\begin{frontmatter}



\title{A covariant simultaneous action for branes}


\author{Riccardo Capovilla and Giovany Cruz}

\address{
Departamento de F\'{\i}sica, Cinvestav-IPN, Av. Instituto Polit\'ecnico Nacional 2508,
col. San Pedro Zacatenco, 07360, Gustavo A. Madero, Ciudad de M\'exico, M\'exico}

\ead{capo@fis.cinvestav.mx, gcruz@fis.cinvestav.mx}

\begin{abstract}
A covariant simultaneous action for branes in an arbitrary curved background spacetime is considered. The action  depends on
a pair of independent field variables, the brane embedding functions, through the canonical momentum of a reparametrization invariant geometric model for the brane, 
and an auxiliary vector field. The form of the action is analogous to a symplectic potential. Extremization of the simultaneous action produces at once the equations of motion and the Jacobi equations for the brane geometric model, and it also  provides a convenient shortcut towards its second variation. 
In this note, we  consider geometric models depending only on the intrinsic geometry of the brane worldvolume, and discuss briefly the generalization to 
extrinsic geometry dependent models. The approach  is illustrated for  Dirac-Nambu-Goto [DNG] branes. For a relativistic particle, a simultaneous
action was introduced by Ba$\dot{\mbox{z}}$a\'{n}ski, that served as an inspiration for the present work. 
\end{abstract}




\end{frontmatter}


Brane mechanics is the study of the dynamics of relativistic extended objects \cite{Carterbrane,defos}, like relativistic 
strings, domain walls, and objects of higher dimension. In a relativistic setting, branes describe 
physical systems with degrees of freedom localized on spacelike  sub-manifolds in an ambient
fixed background spacetime, or ``bulk". The organizing principle of brane mechanics is the symmetry of reparametrization invariance. Branes are described by a reparametrization invariant local action that is a functional of the geometry of the
worldvolume spanned by the brane in its evolution. Reparametrization invariance and the background diffeomorphism invariance, or Poincar\'e invariance for a Minkowski
background, limit strongly the number of possible geometric models available. 
Besides its importance in relativistic mechanics, brane mechanics is of interest in a variety of contexts.    
Brane mechanics plays an essential role  in the framework of braneworlds scenarios, where the four-dimensional universe is considered as a brane embedded in a higher dimensional fixed background, curved or flat, see {\it e.g.} \cite{MK} and references therein, and  in M-theory, where branes are considered as fundamental objects, see {\it e.g.}  \cite{simon}. Other applications
in the realm of astrophysics and black hole physics where physical degrees of freedom are localized on submanifolds of spacetime can also be listed. 
An additional recent motivation comes from the study of entangling surfaces, and the general subject of boundary entropy \cite{armas,netta,speranza}.

The main point of this note is to offer a covariant simultaneous action that depends both on 
the shape functions describing the brane worldvolume and an additional auxiliary vector field. 
We consider a brane described by a reparametrization 
invariant local action, and thus by a geometrical model. 
Variation of the simultaneous action produces at once both the brane
equations of motion,  and the brane Jacobi equation, describing perturbations of the brane
about a solution of its equations of motion. 
The Jacobi equation  can  be considered  also as an equation describing the effect of the background curvature on 
nearby branes in their evolution. In this case, the additional vector field can be interpreted as a deviation vector, as in the 
limit case of a relativistic particle, where it satisfies the familiar geodesic  deviation equation \cite{Wald}. 
In other words, the simultaneous action gives at once the first and second order perturbations of the brane. For the
latter, it can be considered as a convenient shortcut, since it reduces the calculation of the second variation to a first variation.

For a  relativistic 
particle, a simultaneous variational principle was  proposed by Ba$\dot{\mbox{z}}$a\'{n}ski in \cite{bz77b,bz89}, that
gives both the geodesic equation and the geodesic deviation equation. Ba$\dot{\mbox{z}}$a\'{n}ski had previously considered
the case of a general field theory in \cite{Bazpol}, motivated by work in the calculus of variations that goes back to 
Carath\'{e}odhory  \cite{Cara}. 
For the case of a Dirac-Nambu-Goto [DNG] brane this generalization has been considered before by various authors, {\it e.g.} \cite{Roberts,PK},
but the importance of the underlying variational structure and its generality does not appear to have been properly recognized. 

 The generality of the framework is to be intended in the sense that the simultaneous action captures, in a first
 approximation, the effect of the  background spacetime on the evolution of the brane. If the brane is seen
 as a test object, then one obtains information about the background curvature. On the other hand, if one is interested
 in the dynamics  of the brane itself, and its perturbations, the background curvature can be seen as an external force. 
Moreover, the generality is also intended in the sense that there is a  simultaneous action for any type of geometric brane model.
In this note we focus on geometric models that  depend on the worldvolume intrinsic geometry only. For models that depend  also  on the worldvolume extrinsic geometry,
the simultaneous action maintains its form only in a flat background. For a version where the action becomes a free energy describing elementary excitations of lipid membranes   
see \cite{simuhel}.

We start by setting our notation. The brane of dimension $d$ sweeps  in its evolution a worldvolume $w$ of dimension $d+1$. The worldvolume $w$ can be
described in parametric form  by the  (time-like) local embedding functions $x^\mu =  X^\mu (\xi^a)$,
 in a fixed background spacetime 
$\{ M , g_{\mu\nu} \} $ of dimension $N+1$ with local coordinates $x^\mu$, and where $\xi^a$ are local coordinates for $w$. 
 ($\mu, \nu, \dots = 0, 1, \dots, N;  a, b, \dots = 0, 1, \dots, d$). The $d+1$ tangent vectors to the worldvolume  $w$ are
  $X^\mu_a  = \partial_a X^\mu = \partial X^\mu / \partial \xi^a $, and the induced metric, or first fundamental form, on $w$ is
$
\gamma_{ab} = g_{\mu\nu} X^\mu_a X^\nu_b 
$, with determinant $\gamma$.  
Spacetime indices are lowered and raised with the 
spacetime metric $g_{\mu\nu}$ and its inverse $g^{\mu\nu}$. Tangential indices are raised with the inverse induced metric $\gamma^{ab}$ and lowered with $\gamma_{ab}$.
The background spacetime covariant derivative  $\nabla_\mu$ is assumed to be torsionless and compatible with the background metric $g_{\mu\nu}$, with Riemann curvature
$[ \nabla_\mu , \nabla_\nu ] V^\rho =  - R_{\mu\nu\sigma}{}^\rho V^\sigma$, where we follow the conventions of \cite{Wald}. We denote with $\overline{\nabla}_a$ the worldvolume covariant derivative compatible with the induced metric $\gamma_{ab}$,
 $\overline{\nabla}_a \gamma_{bc} = 0$, with Riemann curvature $ {\mathcal R}_{abc}{}^d $. 
The extrinsic curvature tensor, or second fundamental form, is defined as $K_{ab}{}^i = - g_{\mu\nu} n^{\mu\, i} \nabla_a X^\nu_b$, and $K^i = \gamma^{ab} K_{ab}{}^i $ denotes the mean extrinsic curvature.  The spacelike normal vectors $n^{\mu\, i}$  to the
worldvolume  $(i,j \dots = 1, 2, \dots, N-d)$ are defined up to a sign and a rotation by $ g_{\mu\nu}  X^\mu_a n^\nu{}_i = 0$ and 
$g_{\mu\nu} n^\mu{}_i n^\nu{}_j = \delta_{ij}$. Our apologies for the plethora of indices, but we find it clearer than idiosyncratic or abstract notations that emphasize
manifest covariance, at the expense of convenience.

Consider  the general case of a brane  model described by a reparametrization invariant  local action,
\be
S [ X ] = \int_w {\mathcal L} (X^\mu_a )\,,
\label{eq:a}
\ee
where the Lagrangian density of weight one ${\mathcal L} (X^\mu_a)$ depends at most on first derivatives of the embedding functions, {\it i.e.} only on the intrinsic geometry of the worldvolume $w$. 
We have absorbed the differential $d^{d+1} \xi$ in the 
integral sign, henceforth. The simplest example is the DNG model, proportional to the volume of $w$,  with $\mu$ a constant tension,
\be
S_{DNG} [X]  = - \mu \int_w \sqrt{-\gamma}\,.
\label{eq:dng}
\ee
As a second example, the Regge-Teitelboim model for gravity {\it \'{a} la string} \cite{RT}, or `geodetic gravity' \cite{KD}, is defined by the action
\be
S_{RT} [X]  = - \rho \int_w \sqrt{-\gamma} \, {\mathcal R}\,,
\label{eq:rt}
\ee
where $\rho$ is a constant, and ${\mathcal R}$ denotes the scalar curvature of the worldvolume covariant derivative $\overline{\nabla}_a$ compatible with the induced metric $\gamma_{ab}$,
{\it i.e.} $\overline{\nabla}_a \gamma_{bc} = 0$.

The standard variational approach considers the first variation of the action with respect to an infinitesimal variation of the fields, in this case the embedding functions. Its vanishing, under appropriate boundary conditions, produces the equations of motion for the brane. The second variation of the action, assuming that the equations of motion are satisfied,
gives a quadratic form, or index, that establishes the stability of the field configurations.
In addition, the vanishing of the second variation produces the Jacobi equation for the brane, that in the case of the relativistic particle is known as the geodesic deviation
equation. 

The simultaneous action is quite compact and elegant,
\be
S [ X , \eta ] = \int_w \, {\mathcal P}_\mu{}^a  \, \nabla_a \eta^\mu\,.
\label{eq:b}
\ee
The two independent fields are  $X^\mu (\xi^a)$, the worldvolume embedding functions, and an auxiliary vector field $\eta^\mu (\xi^a)$. $\nabla_a$ is the directional covariant derivative along the tangent vectors, $\nabla_a = X^\mu_a \nabla_\mu$.
  Note the analogy with a symplectic potential of the form $ p dq$. 
 The canonical momentum is the worldvolume vector density and spacetime 1-form,
\be
{\mathcal P}_\mu{}^a = {\partial {\mathcal L} \over \partial X^\mu_a}\,.
\label{eq:mom}
\ee  

The symmetries of the simultaneous action are worldvolume diffeomorphisms, or reparametrization invariance, and a constant translation of the auxiliary field. In addition,
the action is a scalar under background diffeomorphisms, or Poincar\'{e} transformations in the case of a flat Minkowski background.

The  first variation of the simultaneous action with respect to an infinitesimal variation of the auxiliary field $\eta^\mu \to \eta^\mu + \delta \eta^\mu$, keeping the embedding functions 
fixed, is  simply
\be
\delta_\eta S [ X, \eta ] |_X =  
 \int_w  {\mathcal P}_\mu{}^a  \nabla_a \delta \eta^\mu = - \int_w \left( \nabla_a {\mathcal P}_\mu{}^a \right) \delta \eta^\mu =  \int_w {\mathcal E}_\mu ({\mathcal L}) \delta \eta^\mu 
 \label{eq:eta1}
 \ee
 where we have integrated by parts,  and we have neglected a boundary term at the moment, for the sake of simplicity.
 Further, we have identified ${\mathcal E}_\mu ({\mathcal L}) = - \nabla_a {\cal P}_\mu{}^a$ as the Euler-Lagrange derivative of the Lagrangian density ${\cal L}$. The vanishing of this first variation under an arbitrary variation of the auxiliary field, $\delta_\eta S [ X, \eta ] |_X = 0$,  therefore implies the equations of motion for the brane
 \be 
 {\cal E}_\mu ({\mathcal L}) = - \nabla_a {\mathcal P}_\mu{}^a= 0\,.
 \label{eq:eom}
 \ee
Note that  this part of the simultaneous variational principle is model independent, in the sense that it does not depend on the specific form of 
${\cal L} (X^\mu_a)$, and that it uses the fact that the equations of motion can be written as a conservation law, \cite{Carterbrane,ACG}. This is 
simply a dimensionally reduced version of the conservation of the stress-energy tensor for a relativistic theory.
 
 Turning to the variation of the simultaneous action with respect to a variation of the 
  embedding functions, $X^\mu \to X^\mu + 
  \delta X^\mu$, we adopt  a covariant variational derivative 
  \be
  \delta_X =  \delta X^\mu \nabla_\mu 
  \ee
   along the variation $\delta X^\mu$, as opposed 
  to  the usual variation that would involve a one-parameter family of embedding functions  $X^\mu (\xi^a , s )$, with $s$ an arbitrary parameter, and a coordinate variation $\delta X^\mu = (\partial X^\mu (\xi^a ,s ) / \partial s)|_{s=0}$
  as  is customary in the calculus of variations \cite{GF}. For a careful treatment of the covariant variational derivative see {\it e.g.} \cite{bz77b,defos}.  The clear advantage is that $\delta_X g_{\mu\nu} = 0$, since the background covariant derivative is metric compatible.

  Keeping the auxiliary field $\eta^\mu$ fixed, we have
  \be
\delta_X S [ X, \eta ] |_\eta  = 
 \int_w [ (\delta_X  {\cal P}_\mu{}^a )  \nabla_a \eta^\mu + {\cal P}_\mu{}^a \delta_X \nabla_a \eta^\mu ]\,.
 \label{eq:delx}
   \ee
 Using the definition of the canonical momentum (\ref{eq:mom}), the first term can be written as 
 \ba
 (\delta_X  {\cal P}_\mu{}^a )  \nabla_a \eta^\mu &=& {\delta^2 {\cal L} \over \partial X^\nu_b \partial X^\mu_a} (\delta_X X^\nu_b ) \nabla_a \eta^\mu \nn \\
 &=& {\cal H}_{\nu\mu}^{ba} (\delta_X X^\nu_b ) \nabla_a \eta^\mu = {\cal H}_{\nu\mu}^{ba} (\nabla_b \delta X^\nu ) \nabla_a \eta^\mu \,,
 \label{eq:hess}
\ea
 where we have defined the Hessian 
 \be
  {\cal H}_{\nu\mu}^{ba}  = {\delta^2 {\cal L} \over \partial X^\nu_b \partial X^\mu_a}\,,
  \label{eq:hess}
 \ee
 and in the second line we have used the fact that variation and partial derivative commute, $\delta_X X^\nu_b = \nabla_b \delta X^\nu$ \cite{defos}.
 Note that the Hessian is degenerate, it admits null eigenvectors, because of the gauge freedom associated with reparametrization invariance.
 The Hessian is symmetric in pairs of indices
 \be
  {\cal H}_{\nu\mu}^{ba} =  {\cal H}_{\mu\nu}^{ab}\,.
  \label{eq:symm}
  \ee
 To obtain the second term in (\ref{eq:delx}),
 one  takes into account the dependance of the covariant derivative
 on the embedding functions, and exploits the  Bianchi identity, together with the independence of the field variables  $X^\mu$ and $\eta^\mu$, in the sense that
 $\delta_X \eta^\mu = 0$, to
obtain
 \be
 {\cal P}_\mu{}^a \delta_X \nabla_a \eta^\mu = {\cal P}_\mu{}^a [ \delta_X, \nabla_a] \eta^\mu = -  R_{\rho\sigma\nu}{}^{\mu}  \delta X^\rho X^\sigma_a  \eta^\nu {\cal P}_\mu{}^a\,, 
 \label{eq:curv}
 \ee

Inserting (\ref{eq:hess}) and (\ref{eq:curv}) in the variation (\ref{eq:delx}) results immediately in
\be
\delta_X S [ X, \eta ] |_\eta  = 
 \int_w \left[  {\cal H}_{\nu\mu}^{ba} (\nabla_b \delta X^\nu ) \nabla_a \eta^\mu -  R_{\rho\sigma\nu}{}^{\mu}  \delta X^\rho X^\sigma_a  \eta^\nu {\cal P}_\mu{}^a \right]\,.
\label{eq:second}
 \ee
At this point, if we identify the auxiliary field with the embedding functions variation, $\eta^\mu = \delta X^\mu$, we have obtained in a few lines a general 
expression for the second variation of the geometric model (\ref{eq:a}), making evident the convenience of the simultaneous action approach. This can be compared with more laborious
approaches available in the literature. Moreover, the central role of the Hessian is brought to the forefront.

To arrive at the Jacobi equations for the brane, all is needed is integration by parts of the first term in  (\ref{eq:second}), neglecting a boundary term. This yields
\be
\delta_X S [ X, \eta ] |_\eta  =  -  \int_w  \left\{ \left[ \nabla_b \left(  {\cal H}_{\nu\mu}^{ba} \nabla_a \eta^\mu \right) \right]  +  R_{\nu \sigma\rho}{}^{\mu}  X^\sigma_a  \eta^\rho {\cal P}_\mu{}^a \right\} \delta X^\nu\,.
\label{eq:jacobi}
\ee
The vanishing of this variation gives the Jacobi equations for the brane
\be
{\cal J}_\nu ({\cal L }) = \nabla_b \left(  {\cal H}_{\nu\mu}^{ba} \nabla_a \eta^\mu \right)  + R_{\nu\sigma\rho}{}^{\mu}   X^\sigma_a  \eta^\rho {\cal P}_\mu{}^a = 0\,.
\ee
The first `kinetic term' involves the Hessian of the geometric model. To obtain this, a specific choice of the geometric model is needed. The second term can be interpreted as providing an `external force' on $\eta^\mu$ caused by a non-trivial background spacetime curvature. To our knowledge, the general structure of this  equation has not been presented
before in the brane mechanics literature.
    
To summarize,  simultaneously  one obtains the brane equations of motion and the Jacobi equations for the brane. 
It is somewhat amusing the interchange of tasks: the variation with respect to the deviation vector gives the  
 equations of motion for the shape functions, and the variation with respect to the shape functions
 gives the Jacobi equations for the deviation vector. The added benefit of the simultaneity is that it is not necessary to impose by hand 
 that the Jacobi equation is to be evaluated on-shell. Simultaneity makes  it automatic.

In order to illustrate the general  formalism, we conside the DNG model (\ref{eq:dng}) in an arbitrary background curved spacetime.  This can be compared 
to the more conventional treatment offered by Carter in a covariant approach \cite{Carter}. The conclusions are the same, but in a different guise.
For a DNG model, we have that the canonical linear momentum is
\be
{\cal P}_\mu{}^a = {\partial {\cal L}_{DNG} \over  \partial X_a^\mu  } =  - \, \mu \sqrt{-\gamma} \, g_{\mu\nu} \gamma^{ab} X^\nu_b \,.
\ee
Note that it is tangential to the worldvolume, as is the case for any intrinsic geometric model, and isotropic.

The simultaneous action  (\ref{eq:b}) takes the form
\be
S_{DNG} [ X , \eta ] = - \mu \int_w \sqrt{-\gamma} \, g_{\mu\nu}  \gamma^{ab} X^\nu_b \nabla_a \eta^\mu\,.
\label{eq:bdng}
\ee

The vanishing of the first variation of this simultaneous action with respect
to $\eta$, see (\ref{eq:eta1}), gives the DNG equations of motion, as in (\ref{eq:eom}),
\be
\nabla_a {\cal P}_\mu{}^a = - \mu \nabla_a \left( \sqrt{-\gamma} \, g_{\mu\nu} \gamma^{ab} X^\nu_b \right)\,,
\ee
Its normal projection gives 
\be
n^\mu{}_i \nabla_a {\cal P}_\mu{}^a = \mu \sqrt{-\gamma} K_i = 0\,,
\ee
{\it i.e.} the vanishing of the worldvolume mean extrinsic curvature $K_i = \gamma^{ab} K_{ab\,i}$. This is the relativistic version of the equilibrium condition for a minimal surface in an arbitrary codimension.

For the variation of the simultaneous action with respect to a variation of the embedding functions, see (\ref{eq:hess}), we need a short calculation to yield  
for the Hessian (\ref{eq:hess})
\be
 {\cal H}_{\nu\mu}^{ba}  
=   \sqrt{-\gamma}   \left(
 \gamma^{ab} n_{\mu\nu} + X_{\mu\nu}^{ab} \right) \,,
 \label{eq:dp}
\ee
where we have used the completeness relationship
\be
g^{\mu\nu} = h^{\mu\nu} + n^{\mu\nu} =  \gamma^{ab} X^\mu_a X^\nu_b + n^\mu{}_i n^{\nu\, i} \,,
\label{eq:comple}
\ee
to define the tangential and normal projectors $h^{\mu\nu}$ and $n^{\mu\nu}$, respectively.
The tangential bivector that appears in (\ref{eq:dp}) is 
\be
X^{\mu\nu}_{ab} = 2X^{[\mu}_a X^{\nu]}_b \,,
 \ee 
with all indices raised with $\gamma^{ab}$ and lowered with $g_{\mu\nu}$, respectively. 
It vanishes identically
for a relativistic particle, and it respects the symmetry property (\ref{eq:symm}).

Inserting (\ref{eq:dp}) for $ {\cal H}_{\nu\mu}^{ba}  $ in the first variation (\ref{eq:second}) gives immediately
  \be
\delta S_X [ X, \eta ] |_\eta  = - \mu 
 \int_w  \sqrt{-\gamma}  \left[ \left(
 \gamma^{ab} n_{\mu\nu} + X_{\mu\nu}^{ab} \right) \left( \nabla_b \delta X^\nu  \right) \nabla_a \eta^\mu -  R_{\rho\sigma\nu}{}^{\mu}   \delta X^\rho h^\sigma{}_\mu  \eta^\nu \right] \,.
 \label{eq:secdng}
 \ee
If at this point we identify the auxiliary field $\eta^\mu$ with the variation of the embedding functions $\delta X^\mu$, this expression reproduces the second variation of the DNG action with respect to variations of
the embedding functions \cite{Carter,BC00}. 
By integrating by parts the first term, and neglecting a boundary term, we obtain
 \be
\delta S_X [ X, \eta ] |_\eta  =  \mu 
 \int_w \left\{ \nabla_b \sqrt{-\gamma}  \left[ \left(
 \gamma^{ab} n_{\mu\nu} + X_{\mu\nu}^{ab} \right)  \nabla_a \eta^\mu \right\} \delta X^\nu  +  \sqrt{- \gamma} R_{\rho\sigma\nu}{}^{\mu}  \delta X^\rho h^\sigma{}_\mu  \eta^\nu \right]  \,.
 \label{eq:secdng}
 \ee
Setting this variation to vanish, $\delta S_X [ X, \eta ] |_\eta = 0$, gives the Jacobi equations for the DNG brane
\be
{\cal J}_\nu ({\cal L_{DNG}}) = \mu  \nabla_b \sqrt{-\gamma}  \left[ \left(
 \gamma^{ab} n_{\mu\nu} + X_{\mu\nu}^{ab} \right)  \nabla_a \eta^\mu \right]  + \mu \sqrt{-\gamma} R_{\nu\sigma\mu}{}^{\rho}   h^\sigma{}_\rho  \eta^\mu  = 0 \,.
 \label{eq:jacd}
\ee
The first kinetic term  differs from the one of the case of a particle, see below, for the presence, in general, of the 
 `friction' term  due to the tangential bivector. It should be noted that these Jacobi equations have been obtained before  in 
 a different, but ultimately equivalent form in \cite{PK}. These equations provide a generalization to branes of the well known geodesic
 deviation equation for particles.

For the special case of a relativistic particle, the degenerate case of a brane of dimension zero, the  simultaneous action was introduced by Ba$\dot{\mbox{z}}$a\'{n}ski  in \cite{bz77b, bz89}, applying a general formalism he developed in \cite{Bazpol}. 
The simultaneous Ba$\dot{\mbox{z}}$a\'{n}ski  action is given by the specialization of the simultaneous DNG action (\ref{eq:bdng}) to a 0-brane as
\be
S [ X , \eta ] = m\int_w  {g_{\mu\nu} U^\nu \over \sqrt{-\gamma}}  \nabla_U \eta^\mu\,,
\ee 
where we set $\sigma = m$. $U^\nu = d X^\nu (\xi ) / d\xi$ is the time-like tangent vector to the particle worldline,
$\gamma = g_{\mu\nu} U^\mu U^\nu$ is the one-dimensional metric, and the directional derivative is  $\nabla_U = U^\mu \nabla_\mu$.
The apparent sign mistake is due to the one-dimensional fact
$\sqrt{- \gamma} \gamma^{-1} = - 1/\sqrt{-\gamma}$, that illustrates how the special case of a worldline can be misleading in a generalization to a higher
dimensional brane. 
Note that the canonical momentum takes the form
 \be
 P_\mu =    {m\over \sqrt{-\gamma}} g_{\mu\nu} U^\nu \,.
\ee

The equations of motion that are obtained by the vanishing of the first variation of this simultaneous action are given by
\begin{eqnarray}
\delta \eta:& \qquad  - m \nabla_U   \left( {1 \over \sqrt{-\gamma}} g_{\mu\nu} U^\mu \right) = 0\,,
  \label{eq:eom1} \\
\delta X:&  \qquad   m \nabla_U \left[ {1 \over  \sqrt{-\gamma}}  (g_{\mu\nu} - h_{\mu\nu})   \nabla_U \eta^\nu \right] + m R_{\mu \sigma \alpha \nu}  U^\sigma
U^\alpha \eta^\nu = 0\,. 
\label{eq:jac1}
\end{eqnarray}
The first is simply the geodesic equation. The
second is the Jacobi equation, or geodesic deviation equation, when $\eta^\mu$ is
 interpreted as the deviation  vector between neighbouring geodesics \cite{Wald}. The Jacobi equation can be rewritten 
 as
 \be
 m \nabla_U \left[ {1 \over  \sqrt{-\gamma}}  n_{\mu\nu}  \nabla_U \eta^\nu \right] + m R_{\mu \sigma \alpha \nu}  U^\sigma
U^\alpha \eta^\nu = 0\,,
 \ee
 using the one-dimensional version of the completeness relationship (\ref{eq:comple}), with the one-dimensional
 particularity  for the tangential projector,  $ h^{\mu\nu} = U^\mu U^\nu / (g_{\rho\sigma} U^\rho U^\sigma )$.
The generalization of this result from the special case of a relativistic particle, to the more general case of a brane
of arbitrary dimension did require a better appreciation of the fundamental role of the canonical momentum. 

For the Regge-Teitelboim model, or `geodetic gravity', the covariant simultaneous action  takes the form
\be
S [ X, \eta ] = \rho \int_w \sqrt{-\gamma} \,  g_{\mu\nu}  {\cal G}^{ab} \, X^\mu_a \nabla_b \eta^\nu\,. 
\ee
where ${\cal G}_{ab} = {\cal R}_{ab} - (1/2) {\cal R} \gamma_{ab}$ denotes the Einstein tensor of the covariant derivative $\overline{\nabla}_a$ compatible with
the induced metric $\gamma_{ab}$. The resulting equations of motion are well known as $ {\cal G}^{ab} K_{ab}{}^i = 0$ \cite{RT,KD}, and they can be derived 
from a variation of this action with respect to the auxiliary field $\eta^\mu$. The Jacobi equations involve a more laborious calculation, that will be presented elsewhere,
together with its physical consequences in brane world scenarios \cite{CG2}. 

In the case of a dependance of the geometric model on the worldvolume extrinsic geometry, the covariant simultaneous action mantains its form
(\ref{eq:b}) in a flat Minkowski background, with the appropriate expression for the canonical momentum for a higher derivative theory, given by
\be
P_\mu{}^a = {\partial {\cal L} \over \partial X^\mu_a} - \nabla_a \left( {\partial {\cal L} \over \partial X^\mu_{ab}} \right)\,,
\ee 
where $X^\mu{}_{ab} = \nabla_a \nabla_b X^\mu$ is the covariant second derivative of the embedding functions, see \cite{simuhel} for a treatment in the context of
the free energy for lipid membranes, where an analogue of the same variational problem occurs.

In conclusion, in this note  we have introduced a covariant simultaneous action that, given any reparametrization invariant local geometric model for a relativistic brane,
produces at once the equations of motion and the Jacobi equations for the model. At the same time, the action provides a convenient shortcut towards the second
variation of the geometric model, as an alternative path towards stability studies. The simultaneous action can be easily extended to accommodate for possible fields
living on the brane, or for `pressure terms' that arise in fundamental branes. Of course, these addictions would gravely affect the simple elegance of the 
covariant simultaneous principle, but would be useful in applications where external forces are relevant.





\section*{Acknowledgements}

GC aknowledges partial support from CONACYT

\end{document}